\documentclass[11pt]{article}
\usepackage[left=1in,top=1in,right=1in,bottom=1in]{geometry}
\usepackage{times}

\usepackage{graphicx}
\usepackage{amsmath}
\usepackage{amsthm}
\usepackage{booktabs}
\usepackage{algorithm}
\usepackage{algorithmic}
\usepackage[switch]{lineno}

\usepackage{amssymb}
\usepackage{verbatim}
\usepackage{url}

\usepackage{siunitx}
\allowdisplaybreaks

\newtheorem{definition}{Definition}
\newcommand{\algorithmicbreak}{\textbf{break}}
\newcommand{\BREAK}{\STATE \algorithmicbreak}

\begin{document}

\title{Computing Evolutionarily Stable Strategies in Imperfect-Information Games}

\author{
Sam Ganzfried\\
Ganzfried Research, Cornell University\\
\texttt{sam.ganzfried@gmail.com}
}

\date{\vspace{-5ex}}

\maketitle

\begin{abstract}
We present an algorithm for computing evolutionarily stable strategies (ESSs) in symmetric two-player perfect-recall extensive-form games of imperfect information. The algorithm operates directly in sequence form, avoiding explicit enumeration of normal-form supports. In a biologically motivated cancer signaling game, the algorithm recovers all seven symmetric Nash equilibria and identifies all three ESSs, each with a clear biological interpretation. Experiments on random imperfect-information games evaluate scalability as the numbers of private signals and available actions increase. 
\end{abstract}

\section{Introduction}
\label{se:intro}
The standard solution concept for evolutionary game theory, which typically models strategic interactions between evolving biological populations, is evolutionarily stable strategy (ESS).
A mixed strategy in a two-player symmetric game is an evolutionarily stable strategy if it is robust to being overtaken by a mutation strategy. Formally, mixed strategy $\mathbf{x}^\star$ is an ESS if for every mixed strategy $\mathbf{x}$ that differs from $\mathbf{x}^\star$, there exists $\varepsilon_0 = \varepsilon_0(\mathbf{x}) > 0$ such that, for all $\varepsilon \in (0,\varepsilon_0)$, 
\begin{equation*}
(1-\varepsilon)u_1(\mathbf{x},\mathbf{x}^\star) + \varepsilon u_1(\mathbf{x},\mathbf{x}) <  (1-\varepsilon)u_1(\mathbf{x}^\star,\mathbf{x}^\star)+\varepsilon u_1(\mathbf{x}^\star,\mathbf{x}).
\label{eq:ess}
\end{equation*}
From a biological perspective, we can interpret $\mathbf{x}^\star$ as a distribution among ``normal'' individuals within a population, and consider a mutation that makes use of strategy $\mathbf{x}$, assuming that the proportion of the mutation in the population is $\varepsilon$. In an ESS, the expected payoff of the mutation is smaller than the expected payoff of a normal individual, and hence the proportion of mutations will decrease and eventually disappear over time, with the composition of the population returning to being mostly $\mathbf{x}^\star$. An ESS is therefore a mixed strategy of the column player that is immune to being overtaken by mutations. ESS was initially proposed by mathematical biologists motivated by applications such as population dynamics (e.g., maintaining robustness to mutations within a population of humans or animals)~\cite{Maynard73:Logic,Maynard82:Evolution}. A common example game is the $2 \times 2$ game where strategies correspond to an ``aggressive'' Hawk or a ``peaceful'' Dove strategy. A paper has recently proposed a similar game in which an aggressive malignant cell competes with a passive normal cell for biological energy, which has applications to cancer eradication~\cite{Dingli09:Cancer}. 

A substantial body of work has explored evolutionary game dynamics in \emph{extensive-form} strategic settings, beginning with the foundational analyses of Selten on evolutionarily stable strategies in multi-stage games with imperfect information. Selten formalized how the ESS concept extends beyond normal form to behavioral strategies in extensive-form games, and showed that mutation stability can depend on sequential information sets and continuation payoffs rather than only aggregate action frequencies~\cite{Selten83:Evolutionary,Selten88:Evolutionary}. Subsequent work by Cressman provided a systematic treatment of evolutionary dynamics in extensive-form games, developing dynamical systems and stability results for behavioral strategies in games with sequential structure, chance nodes, and imperfect information~\cite{Cressman03:Evolutionary}. This line of research established that sequential aspects of play---specifically the information available at each decision point---can fundamentally alter evolutionary stability conditions compared to normal-form representations.

Additional contributions have examined evolutionary dynamics on extensive-form or multi-stage games from both theoretical and computational perspectives. Matros analyzed replicator-type dynamics defined directly on the sequence form of an extensive-form game~\cite{Matros01:Evolutionary}, while Gatti et al.\ studied efficient computation of evolutionary trajectories in EFGs and demonstrated that evolutionary behavior can differ sharply from normal-form reductions even in small games~\cite{Gatti13:Efficient}. Beyond these general frameworks, many evolutionary models naturally incorporate imperfect information through signaling or hidden-type structures. Evolutionarily stable behavior in signaling games has been extensively investigated---for example, J\"ager's ESS conditions for costly signaling equilibria~\cite{Jaeger08:Evolutionarily} and Huttegger's evolutionary analysis of Lewis signaling systems~\cite{Huttegger10:Evolutionary}---all of which rely on imperfect information at the level of types or messages. Related work on Bayesian evolutionary games similarly treats agents' types or preferences as private information, yielding evolutionarily stable outcomes in settings with incomplete information such as cooperation under hidden types~\cite{Akcay12:Evolution} and preference evolution in Bayesian environments~\cite{Alger13:Homo}.

Together, this literature demonstrates both the importance and the difficulty of analyzing evolutionary stability in extensive-form settings: sequential play, hidden information, and structured strategy spaces significantly complicate the evaluation of mutant strategies. However, despite substantial theoretical work, there is currently no general algorithmic framework for computing ESSs in extensive-form games, even though such games arise naturally in evolutionary models of signaling, conflict escalation, communication, sequential cooperation, and biological or ecological interactions. This motivates the present work, which develops the first practical algorithm for identifying evolutionarily stable strategies in general extensive-form games of imperfect information.

While Nash equilibrium is defined for general multiplayer games, ESS is traditionally defined specifically for two-player symmetric games. ESS is a refinement of Nash equilibrium. In particular, if $\mathbf{x}^\star$ is an ESS, then $(\mathbf{x}^\star,\mathbf{x}^\star)$ (i.e., the strategy profile where both players play $\mathbf{x}^\star$) is a (symmetric) Nash equilibrium~\cite{Maschler13:Game}. Of course the converse is not necessarily true (not every symmetric Nash equilibrium is an ESS), or else ESS would be a trivial refinement. In fact, ESS is not guaranteed to exist in games with more than two pure strategies per player (while Nash equilibrium is guaranteed to exist in all finite games). For example, while rock-paper-scissors has a mixed strategy Nash equilibrium (which puts equal weight on all three actions), it has no ESS~\cite{Maschler13:Game} (that work considers a version where payoffs are 1 for a victory, 0 for loss, and $\frac{2}{3}$ for a tie).

There exists a polynomial-time algorithm for computing Nash equilibrium (NE) in two-player zero-sum normal-form games, while for two-player non-zero-sum and multiplayer games computing an NE is PPAD-hard and it is widely conjectured that no efficient (polynomial-time) algorithm exists. However, several algorithms have been devised that perform well in practice. The problem of computing whether a two-player game has an ESS was shown to be both NP-hard and CO-NP hard and also to be contained in $\Sigma ^P _2$ (the class of decision problems that can be solved in nondeterministic polynomial time given access to an NP oracle)~\cite{Etessami08:Computational}. Subsequently it was shown that the exact complexity of this problem is that it is $\Sigma ^P _2$-complete~\cite{Conitzer13:Exact}. Note that this result is for determining whether an ESS exists (as discussed above there exist games which have no ESS), not for the complexity of computing an ESS in games for which one exists. Thus, computing an ESS is significantly more difficult than computing an NE, which is not surprising since it is a refinement. Several approaches have been proposed for computing ESSs in two-player normal-form games~\cite{Haigh75:Game,Abakuks80:Conditions,Broom13:Game-Theoretical,Bomze92:Detecting,Mcnamara97:General}. However, direct normal-form support enumeration can be computationally prohibitive for extensive-form games: the number of pure strategies can be exponential in the size of the game tree, making the number of supports doubly exponential.

\section{Evolutionarily Stable Strategies}
\label{se:ess}
We first review the standard definition of evolutionarily stable strategies for symmetric normal-form games then describe how the definition can be extended to extensive-form games.
A \emph{normal-form game} consists of a finite set of players $N = \{1,\ldots,n\}$, a finite set of pure strategies $S_i$ for each player $i$, and a real-valued utility for each player for each strategy vector (aka \emph{strategy profile}), $u_i : \times_i S_i \rightarrow \mathbb{R}$. In a \emph{symmetric normal-form game}, all strategy spaces $S_i$ are equal and the utility functions satisfy the following symmetry condition: for every player $i \in N$, pure strategy profile $(s_1,\ldots,s_n) \in S^n$, and permutation $\pi$ of the players,
\[
u_i(s_1,\ldots,s_n)
=
u_{\pi(i)}(s_{\pi^{-1}(1)},\ldots,s_{\pi^{-1}(n)}).
\]
This allows us to remove the player index of the utility function and just write 
$u(s_1,\ldots,s_n),$ where it is implied that the utility is for player 1 (we can simply permute the players to obtain the utilities of the other players). We will still write $u_i$ for notational convenience, but note that only a single utility function must be specified which will apply to all players.

Let $\Sigma_i$ denote the set of mixed strategies of player $i$ (probability distributions over elements of $S_i$). If players follow mixed strategy profile
$\mathbf{x} = (\mathbf{x}^{(1)},\ldots,\mathbf{x}^{(n)}),$ where $\mathbf{x}^{(i)} \in \Sigma_i$, the expected payoff to player $i$ is
$$u_i(\mathbf{x}^{(1)},\ldots,\mathbf{x}^{(n)}) = \sum_{s_1,\ldots,s_n \in S} x^{(1)}_{s_1} \cdots x^{(n)}_{s_n}u_i(s_1,\ldots,s_n).$$ 
We write $u_i(\mathbf{x}) = u_i(\mathbf{x}^{(i)},\mathbf{x}^{(-i)})$,
where $\mathbf{x}^{(-i)}$ denotes the vector of strategies of all players except $i$.
If all players follow the same mixed strategy $\mathbf{x}$, then for all players $i$ we have
$$u_i(\mathbf{x}) = u_i(\mathbf{x},\ldots,\mathbf{x}) = \sum_{s_1,\ldots,s_n \in S} x_{s_1} \cdots x_{s_n} u_i(s_1,\ldots,s_n).$$ 

\begin{definition}
\label{de:ne}
A mixed strategy profile $\mathbf{x}^\star$ is a Nash equilibrium if for each player $i \in N$ and for each mixed strategy $\mathbf{x}^{(i)} \in \Sigma_i$:
$u_i(\mathbf{x}^{*(i)},\mathbf{x}^{*(-i)}) \geq u_i(\mathbf{x}^{(i)},\mathbf{x}^{*(-i)}).$
\end{definition}

\begin{definition}
\label{de:ne-sym}
A mixed strategy profile $\mathbf{x}^\star$ in a symmetric normal-form game is a symmetric Nash equilibrium if it is a Nash equilibrium and: 
$\mathbf{x}^{*(1)} = \mathbf{x}^{*(2)} = \cdots = \mathbf{x}^{*(n)}.$
\end{definition}

We now specialize to the classical two-player definition of evolutionary stability, which forms the basis for the extensive-form definition and algorithm developed in this paper. The concept can be extended to multiplayer symmetric games using higher-order stability conditions~\cite{Broom97:Multi}.

\begin{definition}
\label{de:ess}
In a two-player symmetric normal-form game, a mixed strategy $\mathbf{x}^\star \in \Sigma_1$ is \emph{evolutionarily stable} if for each mixed strategy $\mathbf{x} \neq \mathbf{x}^\star$ exactly one of the following conditions holds:
\begin{enumerate}
\item $u_1(\mathbf{x}^\star,\mathbf{x}^\star) > u_1(\mathbf{x},\mathbf{x}^\star)$,
\item $u_1(\mathbf{x}^\star,\mathbf{x}^\star) = u_1(\mathbf{x},\mathbf{x}^\star)$ and $u_1(\mathbf{x}^\star,\mathbf{x}) > u_1(\mathbf{x},\mathbf{x})$.
\end{enumerate}
\end{definition}

It has been proven that every symmetric normal-form game has at least one symmetric Nash equilibrium~\cite{Nash51:Non}. It is clear from Definition~\ref{de:ess} that every evolutionarily stable strategy in a two-player symmetric normal-form game must be a symmetric Nash equilibrium (SNE). Thus, a natural approach for ESS computation is to first compute SNEs and then perform subsequent procedures to determine whether they are ESSs.

Imperfect-information games are modeled using extensive-form game trees, where play proceeds from the root node to a terminal leaf node at which point all players receive payoffs. Each non-terminal node has an associated player (possibly \emph{chance}) that makes the decision at that node. These nodes are partitioned into \emph{information sets}, where the player whose turn it is to move cannot distinguish among the states in the same information set. Therefore, in any given information set, a player must choose actions with the same distribution at each state contained in the information set. If no player forgets information that they previously knew, we say that the game has \emph{perfect recall}. A \emph{behavioral strategy} for player $i,$ $\sigma_i \in \Sigma_i,$ is a function that assigns a probability distribution over all actions at each information set belonging to $i$. 

In order to present our new approach we review the \emph{sequence-form representation} for two-player extensive-form games of imperfect information~\cite{Koller94:Fast}. Rather than operate on the full pure strategy space, which has size exponential in the size of the game tree, the sequence-form works with sequences of actions along trajectories from the root node to leaf nodes. For player 1, the matrix $\mathbf{E}$ is defined where each row corresponds to an information set (including an initial row for the ``empty'' information set), and each column corresponds to an action sequence (including an initial column for the ``empty'' action sequence). In the first row of $\mathbf{E}$ the first element is 1 and all other elements are 0; subsequent rows have -1 for the entries corresponding to the action sequence leading to the root of the information set, and 1 for all actions that can be taken at the information set (and 0 otherwise). Thus $\mathbf{E}$ has dimension $c_1 \times d_1$, where $c_i$ is the number of information sets for player $i$ and $d_i$ is the number of action sequences for player $i$. Matrix $\mathbf{F}$ is defined analogously for player 2. The vector $\mathbf{e}$ is defined to be a column vector of length $c_1$ with 1 in the first position and 0 in other entries, and vector $\mathbf{f}$ is defined with length $c_2$ analogously. The matrix $\mathbf{A}$ has dimension $d_1 \times d_2$. Each entry $A_{ij}$ is the sum, over all terminal histories corresponding to player 1's action sequence $i$ and player 2's action sequence $j$, of player 1's payoff multiplied by the product of chance-move probabilities along that history. The matrix $\mathbf{B}$ of player 2's payoffs is defined analogously. 

For two-player general-sum games, the problem of finding a Nash equilibrium is the feasibility problem of finding
$\mathbf{x},$ $\mathbf{y},$ $\mathbf{p},$ $\mathbf{q}$, such that~\cite{Koller94:Fast}:

\begin{equation} \label{eq-lcp}
\begin{aligned}
-\mathbf{A}\mathbf{y} + \mathbf{E}^\top\mathbf{p} &\ge \mathbf{0} \\
-\mathbf{B}^\top\mathbf{x} + \mathbf{F}^\top\mathbf{q} &\ge \mathbf{0} \\
\mathbf{E}\mathbf{x} &= \mathbf{e} \\
\mathbf{F}\mathbf{y} &= \mathbf{f} \\
\mathbf{x} &\ge \mathbf{0} \\
\mathbf{y} &\ge \mathbf{0} \\
\mathbf{x}^\top(-\mathbf{A}\mathbf{y} + \mathbf{E}^\top\mathbf{p}) &= 0 \\
\mathbf{y}^\top(-\mathbf{B}^\top\mathbf{x} + \mathbf{F}^\top\mathbf{q}) &= 0
\end{aligned}
\end{equation}

The final two constraints are called complementarity slackness conditions, and the full system is known
as a linear complementarity problem (LCP). This LCP can be solved using Lemke's algorithm~\cite{Lemke65:Bimatrix} 
or the related Lemke-Howson algorithm~\cite{Lemke64:Equilibrium}.

We consider two-player extensive-form games that are symmetric in their sequence-form representation. After identifying corresponding action sequences for the two players, their feasible realization-plan sets coincide and their payoff matrices satisfy $\mathbf{B}=\mathbf{A}^\top$. We therefore use common constraints $\mathbf{E}\mathbf{x}=\mathbf{e}$, $\mathbf{x}\ge\mathbf{0}$ for both players, taking $\mathbf{F}=\mathbf{E}$ and $\mathbf{f}=\mathbf{e}$. These conditions express symmetry of the strategy spaces and expected payoffs and accommodate simultaneous decisions represented sequentially using information sets. Let $\boldsymbol{\sigma}$ denote a behavioral strategy, and let $u(\boldsymbol{\sigma},\boldsymbol{\tau})$ be the expected payoff of playing $\boldsymbol{\sigma}$ against an opponent playing $\boldsymbol{\tau}$.

We define evolutionary stability over realization plans, identifying behavioral strategies that induce the same realization plan. Accordingly, distinct strategies in the following definition mean strategies with distinct realization plans.

\begin{definition}
\label{de:ess-efg}
A behavioral strategy $\boldsymbol{\sigma}^\star$ is an
\emph{evolutionarily stable strategy} in a symmetric perfect-recall
extensive-form imperfect-information game if for every behavioral
strategy $\boldsymbol{\sigma}$ whose realization plan differs from
that of $\boldsymbol{\sigma}^\star$, exactly one of the following holds:
\begin{enumerate}
    \item $u(\boldsymbol{\sigma}^\star,\boldsymbol{\sigma}^\star)
           > u(\boldsymbol{\sigma},\boldsymbol{\sigma}^\star)$, or
    \item $u(\boldsymbol{\sigma}^\star,\boldsymbol{\sigma}^\star)
           = u(\boldsymbol{\sigma},\boldsymbol{\sigma}^\star)$ and
           $u(\boldsymbol{\sigma}^\star,\boldsymbol{\sigma})
           > u(\boldsymbol{\sigma},\boldsymbol{\sigma})$.
\end{enumerate}
\end{definition}

Evolutionary stability therefore requires that $\boldsymbol{\sigma}^\star$
perform strictly better than any mutant strategy $\boldsymbol{\sigma}$
when both are played against the incumbent population, and—when the two
strategies perform equally well against the incumbent—that
$\boldsymbol{\sigma}^\star$ outperform $\boldsymbol{\sigma}$ in direct
competition. In the sequence-form representation, any candidate
behavioral strategy must satisfy the linear constraints
$\mathbf{E}\mathbf{x} = \mathbf{e}$ and $\mathbf{x} \ge \mathbf{0}$,
ensuring that evolutionary comparisons are carried out over feasible
realization plans within the extensive-form structure.

Our algorithm will focus on the setting of two-player symmetric 
perfect-recall extensive-form games. We note that sequence-form 
representation extends naturally to $n$-player games, as do the definitions
of symmetric games and evolutionarily stable strategies. We will briefly
discuss how our algorithm can be extended to apply to $n$ players for $n > 2.$

\section{Algorithm}
\label{se:algo}
Our main algorithm for ESS computation in two-player symmetric perfect-recall extensive-form games
is given in Algorithm~\ref{al:ess}. Our method iteratively searches for a new SNE $\mathbf{x}$ and tests whether $\mathbf{x}$ is an ESS. We can halt the algorithm as soon as one ESS is found, or continue it to find as many ESSs as possible until a specified termination criterion is met. The algorithm takes the sequence-form game matrices $(\mathbf{E}, \mathbf{e}, \mathbf{A})$ as input, and uses several hyperparameters $(\delta, \varepsilon_p, \varepsilon_s)$ for numerical precision. The values used in the experiments are given in Table~\ref{ta:parameters}.

Algorithm~\ref{al:ess} first computes an initial symmetric Nash equilibrium $\mathbf{x}$ by running the procedure given in Algorithm~\ref{al:sne-initial}. From Nash's theorem we know that a symmetric Nash equilibrium always exists. We then add $\mathbf{x}$ to the list of known SNEs and calculate the expected payoff of playing $\mathbf{x}$ against itself: $v^\star = \mathbf{x}^\top \mathbf{A} \mathbf{x}.$ We then run Algorithm~\ref{al:ess-test} to test whether $\mathbf{x}$ satisfies the ESS conditions. We then run Algorithm~\ref{al:sne-new} to compute a new undiscovered SNE (if one exists), and repeat this procedure until a desired termination criterion is met.

\begin{algorithm}[!ht]
\caption{ESS search in symmetric two-player extensive-form game of imperfect information}
\label{al:ess}
\begin{algorithmic}[1]
\REQUIRE Sequence-form game matrices $(\mathbf{E}, \mathbf{e}, \mathbf{A})$, parameters $\delta > 0$, $\varepsilon_p > 0$, $\varepsilon_s > 0$, termination criterion
\ENSURE List $\mathrm{ESS\_list}$
\STATE $\mathrm{ESS\_list} \gets \emptyset$
\STATE $\mathrm{SNE\_list} \gets \emptyset$
\STATE $\mathbf{x} \gets$ \textsc{INITIAL\_SNE}$(\mathbf{E}, \mathbf{e}, \mathbf{A})$
\STATE $\mathrm{SNE\_list} \gets \mathrm{SNE\_list} \cup \{\mathbf{x}\}$
\STATE $v^\star \gets \mathbf{x}^\top \mathbf{A} \mathbf{x}$
\STATE $(\text{status},F^\star) \gets$ \textsc{ESS\_TEST}$(\mathbf{E}, \mathbf{e}, \mathbf{A}, \mathbf{x}, v^\star, \delta)$
\IF{$\text{status}=\text{INFEASIBLE}$ \OR $F^\star > \varepsilon_p$} 
		\STATE $\mathrm{ESS\_list} \gets \mathrm{ESS\_list} \cup \{\mathbf{x}\}$
\ENDIF
\WHILE{termination criterion not met}
		\STATE $(\mathbf{x}, t^\star) \gets \textsc{NEW\_SNE}(\mathbf{E}, \mathbf{e}, \mathbf{A}, \mathrm{SNE\_list})$ 
    \IF{$t^\star < \varepsilon_s$}
				\BREAK
		\ENDIF
    \STATE $\mathrm{SNE\_list} \gets \mathrm{SNE\_list} \cup \{\mathbf{x}\}$
    \STATE $v^\star \gets \mathbf{x}^\top \mathbf{A} \mathbf{x}$
		\STATE $(\text{status},F^\star) \gets$ \textsc{ESS\_TEST}$(\mathbf{E}, \mathbf{e}, \mathbf{A}, \mathbf{x}, v^\star, \delta)$
		\IF{$\text{status}=\text{INFEASIBLE}$ \OR $F^\star > \varepsilon_p$} 
				\STATE $\mathrm{ESS\_list} \gets \mathrm{ESS\_list} \cup \{\mathbf{x}\}$
		\ENDIF
\ENDWHILE
\STATE \textbf{return} $\mathrm{ESS\_list}$
\end{algorithmic}
\end{algorithm}

Algorithm~\ref{al:sne-initial} computes the initial SNE by solving a quadratically-constrained feasibility program that is a simpler version of the LCP for computing a Nash equilibrium in general (potentially asymmetric) games given by equation system (\ref{eq-lcp}). Algorithm~\ref{al:sne-new} attempts to compute a subsequent SNE given a list of the SNEs that were previously obtained. Let $\{\mathbf{x}^{(1)}, \ldots, \mathbf{x}^{(T)}\}$ denote current list of SNEs. The algorithm finds an SNE that satisfies the same constraints as those in Algorithm~\ref{al:sne-initial} and maximizes the minimum (squared) $L_2$ distance
from an existing SNE. It does this by maximizing a new scalar variable $t$ with additional constraints
$$\|\mathbf{x} - \mathbf{x}^{(j)}\|_2^2 \ge t \mbox{ for all } j = 1,\ldots,T.$$
This quadratically-constrained program is always feasible, since any previously found SNE $\mathbf{x}^{(i)}$, together with $t=0$, provides a feasible solution.
Algorithm~\ref{al:ess} then tests whether the newly found candidate SNE $\mathbf{x}$ is numerically distinct
from the existing SNEs by comparing $t^\star$ to the numerical separation parameter $\varepsilon_s.$ If $t^\star < \varepsilon_s$ then we conclude that no additional 
SNEs exist that are numerically distinct from the existing SNE set, and halt the algorithm. 

\begin{algorithm}[!ht]
\caption{\textsc{INITIAL\_SNE}$(\mathbf{E}, \mathbf{e}, \mathbf{A})$}
\label{al:sne-initial}
\begin{algorithmic}[1]
\STATE Find $\mathbf{x}, \mathbf{p}, \mathbf{s}$ such that:
\STATE \hspace{1em} $\mathbf{E}\mathbf{x} = \mathbf{e}$, \quad $\mathbf{x} \ge \mathbf{0}$ 
\STATE \hspace{1em} $\mathbf{E}^\top \mathbf{p} - \mathbf{A}\mathbf{x} - \mathbf{s} = \mathbf{0}$, \quad $\mathbf{s} \ge \mathbf{0}$ 
\STATE \hspace{1em} $x_i s_i = 0$ for all $i$ 
\STATE \textbf{return} $\mathbf{x}$
\end{algorithmic}
\end{algorithm}

\begin{algorithm}[!ht]
\caption{\textsc{NEW\_SNE} $(\mathbf{E}, \mathbf{e}, \mathbf{A}, \mathrm{SNE\_list})$}
\label{al:sne-new}
\begin{algorithmic}[1]
\REQUIRE $(\mathbf{E}, \mathbf{e}, \mathbf{A})$, $\mathrm{SNE\_list} = \{\mathbf{x}^{(1)}, \ldots, \mathbf{x}^{(T)}\}$
\STATE Maximize $t$ over variables $\mathbf{x}, \mathbf{p}, \mathbf{s}, t$ subject to:
\STATE \hspace{1em} $\mathbf{E}\mathbf{x} = \mathbf{e}$, \quad $\mathbf{x} \ge \mathbf{0}$
\STATE \hspace{1em} $\mathbf{E}^\top \mathbf{p} - \mathbf{A}\mathbf{x} - \mathbf{s} = \mathbf{0}$, \quad $\mathbf{s} \ge \mathbf{0}$
\STATE \hspace{1em} $x_i s_i = 0$ for all $i$
\STATE \hspace{1em} $\|\mathbf{x} - \mathbf{x}^{(j)}\|_2^2 \ge t$ for all $j = 1,\ldots,T$
\STATE $(\mathbf{x}_{\text{new}}, t^\star) \gets$ optimal solution
\RETURN $(\mathbf{x}_{\text{new}}, t^\star)$
\end{algorithmic}
\end{algorithm}

Given a newly found SNE $\mathbf{x}$ and its payoff against itself $v^\star$, Algorithm~\ref{al:ess-test} tests whether $\mathbf{x}$ is in fact an ESS. Since $\mathbf{x}$ is an SNE, we know that $\mathbf{x}^\top \mathbf{A} \mathbf{x} \geq \mathbf{y}^\top \mathbf{A} \mathbf{x}$ for all 
realization sequences $\mathbf{y}.$ We now need to test whether any mutants $\mathbf{y}$ exist that violate both conditions of Definition~\ref{de:ess-efg}. Note that
we cannot have $\mathbf{y}^\top \mathbf{A} \mathbf{x} > \mathbf{x}^\top \mathbf{A} \mathbf{x}$ since that would violate the condition that $\mathbf{x}$ is an SNE. If
$\mathbf{x}^\top \mathbf{A} \mathbf{x} > \mathbf{y}^\top \mathbf{A} \mathbf{x}$, then the mutant fails to violate the first condition. So the only way that a mutant can successfully violate both conditions is if both $\mathbf{x}^\top \mathbf{A} \mathbf{x} = \mathbf{y}^\top \mathbf{A} \mathbf{x}$ and $\mathbf{x}^\top \mathbf{A} \mathbf{y} \leq \mathbf{y}^\top \mathbf{A} \mathbf{y}.$ To test this, we will minimize the objective 
$F(\mathbf{y}) = \mathbf{x}^\top \mathbf{A} \mathbf{y} - \mathbf{y}^\top \mathbf{A} \mathbf{y},$
subject to the constraint
$\mathbf{y}^\top \mathbf{A} \mathbf{x} = v^\star,$
where we have already computed $v^\star = \mathbf{x}^\top \mathbf{A} \mathbf{x}.$
To ensure that $\mathbf{y}$ is numerically distinct we impose the additional constraint $\|\mathbf{y} - \mathbf{x}\|_2^2 \ge \delta^2$. 
If the optimal objective value $F^\star$ is less than or equal to zero then there exists a mutant that violates the conditions and $\mathbf{x}$ is not an ESS:
otherwise, no mutant at distance at least $\delta$ from $\mathbf{x}$ violates the ESS conditions.
To test this in practice numerically Algorithm~\ref{al:ess} compares $F^\star$ to $\varepsilon_p.$
If the problem is infeasible, then no best-response mutant lies at distance at least $\delta$ from $\mathbf{x}$. Algorithm~\ref{al:ess} 
also accepts $\mathbf{x}$ as an ESS in this case.

\begin{algorithm}[!ht]
\caption{\textsc{ESS\_TEST}$(\mathbf{E}, \mathbf{e}, \mathbf{A}, \mathbf{x}, v^\star, \delta)$}
\label{al:ess-test}
\begin{algorithmic}[1]
\STATE Minimize $F(\mathbf{y}) = \mathbf{x}^\top \mathbf{A} \mathbf{y} - \mathbf{y}^\top \mathbf{A} \mathbf{y}$ subject to:
\STATE \hspace{1em} $\mathbf{E}\mathbf{y} = \mathbf{e}$, \quad $\mathbf{y} \ge \mathbf{0}$
\STATE \hspace{1em} $\mathbf{y}^\top \mathbf{A} \mathbf{x} = v^\star$ 
\STATE \hspace{1em} $\|\mathbf{y} - \mathbf{x}\|_2^2 \ge \delta^2$
\IF{infeasible}
		\RETURN (INFEASIBLE, $\bot$)
\ENDIF
\STATE $F^\star \gets$ optimal objective value
\RETURN (FEASIBLE, $F^\star$)
\end{algorithmic}
\end{algorithm}

\begin{table}[!ht]
\centering
\caption{Parameter values used in the algorithm.}
\label{ta:parameters}
\begin{tabular}{|*{2}{c|}} \hline
Parameter &Value\\ \hline
$\varepsilon_s$ & $10^{-3}$\\  \hline 
$\varepsilon_p$ & $10^{-5}$ \\ \hline
$\delta$    & $10^{-2}$ \\  
\hline
\end{tabular}
\end{table}

\section{Experiments}
\label{se:exp}
To solve the subproblems given in Algorithms~\ref{al:sne-initial}, \ref{al:sne-new}, and \ref{al:ess-test} numerically, we used the nonconvex quadratic solver from Gurobi version 13.0.0, which guarantees global optimality~\cite{Gurobi25:Gurobi}, with Java version 14.0.2. The problem in Algorithms~\ref{al:sne-initial} is a quadratically-constrained feasibility program (QCFP), the problem in Algorithm~\ref{al:sne-new} is a quadratically constrained program (QCP), and the problem in Algorithm~\ref{al:ess-test} is a quadratically-constrained quadratic program (QCQP), all of which are nonconvex. We used an Intel Core i7-1065G7 processor with base clock speed of 1.30 GHz (and maximum turbo boost speed of up to 3.9 GHz) with 16 GB of RAM under 64-bit Windows 
11 (8 logical cores/threads). We used the numerical parameter values given in Table~\ref{ta:parameters}. We used the default feasibility tolerance in Gurobi of $10^{-6}$ (meaning that it is possible that constraints are violated by up to $10^{-6}$). Thus our numerical parameters $\varepsilon_s$, $\varepsilon_p$, and $\delta^2$ are all well above this value. We also performed a numerical procedure to the SNE candidates $\mathbf{x}$ returned by Algorithms~\ref{al:sne-initial} and \ref{al:sne-new} that set all action sequence values below $\varepsilon_{CLIP} = 10^{-5}$ to zero and renormalized. For the termination criterion we halt the algorithm if $t^\star < \varepsilon_s.$

We first evaluate our algorithm on a biologically motivated two-player symmetric imperfect-information game that models clonal interaction under uncertain therapy conditions. At the root, a chance node selects the systemic therapy intensity $T \in \{\text{Low},\text{High}\}$, which is not observed by the players. Each intensity is selected with probability 0.5. Each clone $i \in \{1,2\}$ then receives a private noisy signal $S_i \in \{\text{fav},\text{unfav}\}$, where favorable signals are more likely when therapy is low and unfavorable signals are more likely when therapy is high. Specifically, we assume that each player receives a favorable signal with probability 0.8 when therapy is low and with probability 0.2 when therapy is high. Signals are independent across players conditional on $T$, and each clone observes only its own signal, producing genuine imperfect information. After receiving its signal, each player chooses a phenotype from $\{P,R,Q\}$, representing proliferative (therapy-sensitive), resistant-proliferative, and quiescent (dormant) programs commonly studied in tumor ecology.

\begin{figure}[!ht]
\centering
\includegraphics[width=0.9\textwidth,height=0.9\textheight,keepaspectratio]{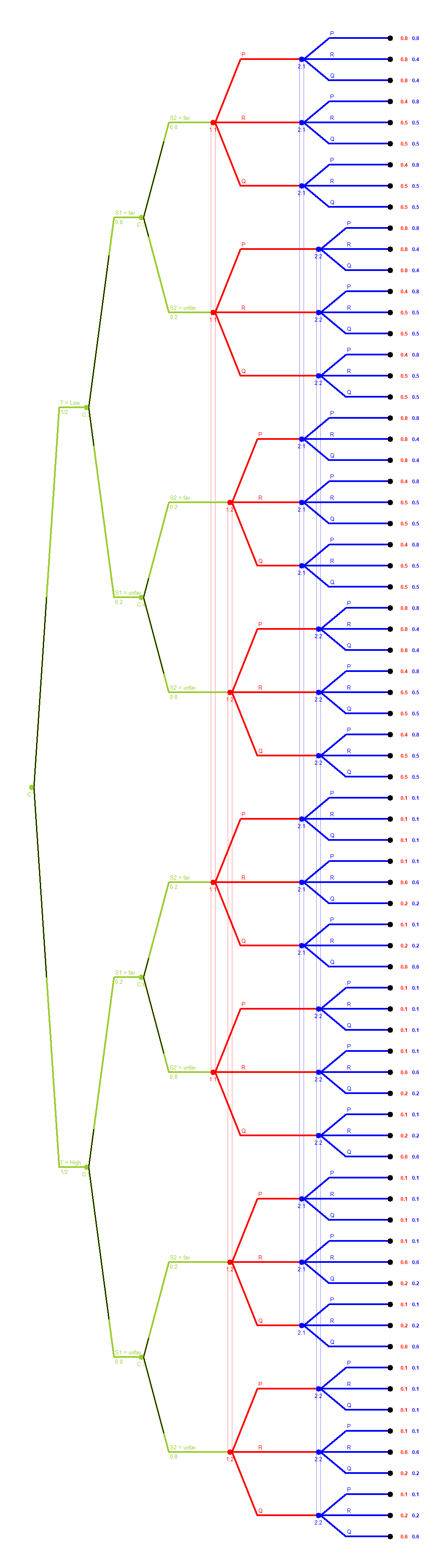}
\caption{Extensive-form tree structure of the cancer signaling game. Nature selects a therapy intensity, then players receive private noisy signals and choose phenotypes at their information sets.}
\label{fi:cancer}
\end{figure}

Payoffs depend on the true therapy level and the phenotypes selected: proliferation $P$ is most advantageous under low therapy, whereas resistance $R$ and quiescence $Q$ yield higher survival under high therapy. Each player selects one of these three phenotypes after observing either a favorable or unfavorable signal. The payoffs under low therapy are given by $\mathbf{A}^{\text{low}}$. We interpret these matrices as follows. If player 1 selects row $i$ and player 2 selects column $j$, then player 1 receives payoff $A^{\text{low}}_{ij}$ and player 2 receives $A^{\text{low}}_{ji}$. Similarly, the payoffs under high therapy are given by $\mathbf{A}^{\text{high}}$. Note that these differ from the final sequence-form matrix $\mathbf{A}$ and are presented to simplify the description of the game. While this game is relatively small it captures key biological features such as environmental uncertainty, private information, and phenotypic plasticity. To give an idea of the game's size and sequential structure the full game tree is given in Figure~\ref{fi:cancer}.

\[
\mathbf{A}^{\text{low}}=
\begin{array}{c|ccc}
 &P&R&Q\\ \hline
P&0.8&0.8&0.8\\
R&0.4&0.5&0.5\\
Q&0.4&0.5&0.5
\end{array}
\qquad
\mathbf{A}^{\text{high}}=
\begin{array}{c|ccc}
 &P&R&Q\\ \hline
P&0.1&0.1&0.1\\
R&0.1&0.6&0.2\\
Q&0.1&0.2&0.6
\end{array}
\]

Our algorithm identifies seven symmetric Nash equilibria. In every equilibrium, players assign probability~1 to the proliferative phenotype~P at the ``fav'' information set, which reflects that proliferation is optimal when the signal indicates likely low therapy. The equilibria differ only in their behavior at the ``unfav'' information set. Here the QCP in Algorithm~\ref{al:sne-new} uncovers a mixture of pure and mixed SNEs that distribute probability across $\{P,R,Q\}$ following an unfavorable signal. At the mixed equilibria, the phenotypes assigned positive probability have equal expected payoffs against the resident strategy, yielding a continuum of best responses. Nevertheless, the seven SNEs listed in Table~\ref{ta:cancer} are isolated. Table~\ref{ta:cancer} summarizes all seven SNEs. The full computation required only 0.287 seconds.

\begin{table}[!ht]
\centering
\caption{SNEs and ESSs in cancer signaling game. All SNEs play $P$ at the ``fav'' information set
and differ only in their response to ``unfav''.}
\label{ta:cancer}
\begin{tabular}{lcc}
\toprule
Unfavorable phenotype & SNE & ESS \\ 
\midrule
$P$                     & \checkmark & \checkmark \\
$R$                     & \checkmark & \checkmark \\
$Q$                     & \checkmark & \checkmark \\
$0.5R+0.5Q$             & \checkmark & $\times$ \\
$0.592P+0.204R+0.204Q$  & \checkmark & $\times$ \\
$0.753P+0.247R$         & \checkmark & $\times$ \\
$0.753P+0.247Q$         & \checkmark & $\times$ \\
\bottomrule
\end{tabular}
\end{table}

Applying the ESS test (Algorithm~\ref{al:ess-test}), only three SNEs are evolutionarily stable. These are exactly the realization plans that (i) play $P$ after a favorable signal and 
(ii) commit to a single pure phenotype $\{P,R,Q\}$ after an unfavorable signal. All four mixed SNEs fail condition~(2) of Definition~\ref{de:ess-efg}: for each resident strategy $\mathbf{x}$, there exists a mutant $\mathbf{y}$ satisfying $u(\mathbf{y},\mathbf{x})=u(\mathbf{x},\mathbf{x})$ but $u(\mathbf{y},\mathbf{y})>u(\mathbf{x},\mathbf{y})$. Such a mutant plays $P$ after a favorable signal and, after an unfavorable signal, plays $R$ for the $R$--$Q$ mixed equilibrium and $P$ for each of the other three mixed equilibria. 
Biologically, the model therefore supports exactly three robust 
clonal response programs: always proliferate; proliferate on favorable signals and 
switch to resistant on unfavorable signals; or proliferate on favorable signals 
and switch to quiescent dormancy on unfavorable signals. In contrast, fine-grained 
mixtures among $\{P,R,Q\}$ are not evolutionarily stable and are susceptible to invasion.

To evaluate the algorithm's scalability we generalized the cancer signaling game by
increasing the number of private signals $S$ and actions $A$ for each player, keeping the
number of hidden states $K$ fixed at 2. For each setting of $S \in \{2,3,4,5\}$ and $A \in \{3,4,5,6\}$ we generated 25 random game instances. These games all have the same sequential structure as the original cancer signaling game, which is encoded in the sequence-form matrix $\mathbf{E}$ and vector $\mathbf{e}.$ (The cancer signaling game has $S = 2$, $A = 3$.) 
For each instance and each hidden state, we independently sample $S$ weights uniformly from $[0,1)$ and normalize them to obtain the conditional signal distribution. Both players use this distribution, with signals drawn independently conditional on the hidden state. The two hidden states have equal prior probability. We set each entry
of the $\mathbf{A}^{\text{high}}$ and $\mathbf{A}^{\text{low}}$ matrices to be uniform random in [-1,1]. We then create the sequence form payoff matrix $\mathbf{A}$ from these matrices. To reduce the number of degenerate games with a continuum of SNEs we add a small amount of noise uniform in $[-\varepsilon_{N},\varepsilon_{N}]$ to entries of $\mathbf{A}$, using $\varepsilon_{N} = 10^{-4}.$ These games form a family of random symmetric extensive-form games of imperfect information with perfect recall parametrized by $A$ and $S.$

The results of our experiments are given in Table~\ref{ta:random}. For each combination of $(S,A)$, we report the mean and median running times as well as the number of SNEs found, the number of ESSs found, and the number of game instances which contained at least one ESS. For the largest games class $S = 5,$ $A = 6$, the median running time was 21.1 seconds, which demonstrates that the algorithm can solve moderate-sized games quickly on a laptop. Mean runtimes substantially exceeded the medians for several settings, particularly the larger games. We report both statistics to distinguish typical runtimes from the influence of slower instances on the averages.

\begin{table}[!ht]
\centering
\caption{Results for random symmetric imperfect-information extensive-form games ($K=2$ hidden states, 25 instances per setting).}
\label{ta:random}
\begin{tabular}{ccccccc}
\toprule
$S$ & $A$ &
Mean Time (s) & Median Time (s) &
Mean \#SNEs (total) &
Mean \#ESSs (total) &
Frac.\ $\ge 1$ ESS \\
\midrule
2 & 3 & 0.241 & 0.098 & 2.40 (60) & 1.20 (30) & 0.80 \\
2 & 4 & 0.368 & 0.246 & 2.64 (66) & 1.36 (34) & 0.84 \\
2 & 5 & 0.491 & 0.388 & 3.08 (77) & 1.32 (33) & 0.76 \\
2 & 6 & 4.152 & 1.990 & 8.76 (219) & 2.32 (58) & 0.80 \\
\midrule
3 & 3 & 0.360 & 0.147 & 2.16 (54) & 1.12 (28) & 0.76 \\
3 & 4 & 1.384 & 0.737 & 5.24 (131) & 1.44 (36) & 0.76 \\
3 & 5 & 1.448 & 1.115 & 3.92 (98) & 1.16 (29) & 0.64 \\
3 & 6 & 10.855 & 4.194 & 8.40 (210) & 1.64 (41) & 0.76 \\
\midrule
4 & 3 & 0.806 & 0.359 & 3.64 (91) & 1.60 (40) & 0.68 \\
4 & 4 & 4.243 & 0.699 & 3.04 (76) & 1.00 (25) & 0.72 \\
4 & 5 & 8.964 & 5.311 & 6.24 (156) & 1.28 (32) & 0.68 \\
4 & 6 & 61.798 & 10.985 & 10.52 (263) & 1.04 (26) & 0.64 \\
\midrule
5 & 3 & 3.791 & 0.676 & 2.52 (63) & 0.76 (19) & 0.52 \\
5 & 4 & 4.106 & 0.943 & 3.44 (86) & 0.92 (23) & 0.64 \\
5 & 5 & 14.521 & 5.338 & 4.52 (113) & 0.76 (19) & 0.44 \\
5 & 6 & 165.155 & 21.145 & 11.56 (289) & 0.96 (24) & 0.44 \\
\bottomrule
\end{tabular}
\end{table}

The results in Table~\ref{ta:random} show that increasing game size generally increases the number of SNEs found. Across all tested settings, the mean number of ESSs found was smaller than the mean number of SNEs. For the largest setting, the algorithm identified an average of 11.56 SNEs and 0.96 ESSs per instance. These findings suggest that evolutionary stability substantially narrows the set of candidate equilibria in the tested games, underscoring the value of computational methods for identifying ESSs in imperfect-information settings.

\section{Conclusion}

We presented a computational framework for identifying evolutionarily stable strategies in two-player symmetric extensive-form games. The approach encodes the game exactly in sequence form, characterizes symmetric Nash equilibria through a feasibility program, then evaluates each candidate via a nonconvex quadratic program that searches for profitable mutant deviations subject to a minimum separation constraint. 

Our experiments demonstrate that the method is practical even for moderately sized imperfect-information games. In a biologically motivated cancer signaling game, the algorithm recovered all seven SNEs and identified all three ESSs, each with a clear biological interpretation. In the random game experiments, which span a range of signal and action complexities, we found that the number of SNEs tends to grow with game size while ESSs remain comparatively rare. The computational burden was modest for many instances: median runtimes were below one second for several small and medium settings, while the largest setting, with $S=5$ signals and $A=6$ actions, had a median runtime of 21.1 seconds. 

This work suggests several promising directions for future research. On the theoretical side, it would be useful to investigate structural properties of extensive-form games that govern the number or stability of SNEs, or that predict when ESSs are likely to exist. On the computational side, improvements in nonlinear optimization or tailored branching strategies may further accelerate the mutant-resistance check. Biologically, the framework opens opportunities to model and analyze evolutionary stability in settings such as tumor–immune interactions, cooperative and competitive signaling among heterogeneous cancer cell subpopulations, treatment-driven evolutionary bottlenecks, and multi-phenotype ecological games where strategic behavior unfolds over latent or partially observed states. Integrating such models with empirical data may ultimately support the design of therapies that exploit evolutionary vulnerabilities in cancer ecosystems.

Our algorithm can be halted early to output the current set of ESSs (or SNEs) that have been discovered. The algorithm returns candidates that pass its numerical ESS test at the chosen separation and payoff thresholds. While we used a specific termination criterion in our experiments of halting the algorithm when no additional SNEs exist with $L_2$ distance from discovered SNEs exceeding a certain threshold, many other termination criteria may be applied. For example, the algorithm could be halted after the first ESS is discovered, after a certain number of SNEs have been found, or after a time limit is reached. Future work can explore improved techniques for addressing degeneracy.

Our framework could potentially be extended to symmetric games with more than two players. In the SNE constraints, expected-payoff expressions would involve products of realization-plan variables from multiple opponents. These products can be represented using auxiliary variables and quadratic equalities, following ideas used in a recent approach to Nash equilibrium computation~\cite{Ganzfried26:Quadratic}. However, extending the ESS verification step would additionally require higher-order stability conditions analogous to those used in multiplayer normal-form games~\cite{Ganzfried25:Computing-Multiplayer}. Developing the corresponding sequence-form optimization formulations and evaluating their computational performance remain directions for future work.

Our algorithm differs from classical support-enumeration approaches for computing ESSs in two-player normal-form games. These approaches examine candidate supports, seeking an SNE with each support and testing whether it is evolutionarily stable. Although support enumeration can be practical for relatively small normal-form games, the number of pure strategies in an extensive-form game can be exponential in the size of the game tree, making exhaustive enumeration of normal-form supports doubly exponential. Our method searches for candidate equilibria directly in sequence form, avoiding this explicit enumeration. Related work on normal-form games uses preprocessing procedures, including a strict-equilibrium shortcut and a pure-mutant screen, to reduce the number of candidates requiring a full ESS test~\cite{Ganzfried25:Computing-Multiplayer}. Such ideas also suggest opportunities for improving our approach. In particular, whether a candidate SNE has a unique best-response realization plan can be tested using linear programming in sequence form; uniqueness establishes evolutionary stability against all distinct realization plans. Developing and evaluating this shortcut and other preprocessing procedures could reduce the number of QCQP solves required by Algorithm~\ref{al:ess-test} and improve the scalability of ESS computation in extensive-form games.

\bibliographystyle{plain}
\bibliography{C://FromBackup/Research/refs/dairefs}

\end{document}